\documentclass[aps,prb,twocolumn,groupedaddress,showkeys]{revtex4}
\usepackage{graphicx}
\begin{document}


\title{An extension of the Hirsch index: Indexing scientific topics and compounds}

\author{M. G. Banks}
\email{m.banks@fkf.mpg.de}
\affiliation{Max-Planck-Institute f\"{u}r Festk\"{o}rperforschung,
Heisenbergstrasse 1, 70569 Stuttgart, Germany}


\date{\today}

\begin{abstract}

An interesting twist of the Hirsch index is given, in terms of an
index for topics and compounds. By comparing both the \textit{hb}
index and \textit{m} for a number of compounds and topics, it can
be used to differentiate between a new so-called hot topic with
older topics. This quick method is shown to help new comers to
identify how much interest and work has already been achieved in
their chosen area of research.

\end{abstract}


\maketitle


The Hirsch index ``\textit{h}
index,"\cite{jeh:hirsch,jeh:preprint} has recently been of much
scientific discussion for its use in ranking the output by a given
scientist. Using the Science Citation Index (SCI) under the Web of
Science (WoS) offered by Thomson ISI (available at
http://isiknowledge.com), and by sorting the authors papers by
ascending order in terms of citations. The \textit{h} index of a
given author is when \textit{h} of his or hers \textit{N}$_{p}$
papers have at least \textit{h} citations each. The corresponding
remaining papers have thus $\leq$ \textit{h} citations each. In
the case where the author has a unique name, this procedure is a
very easy and fast way in order to compare scientists. More
importantly, it does not fully depend on the total number of
papers, \textit{N}$_{p}$, which could be bolstered by conference
proceedings, or the total number of citations or significant
papers. The Hirsch index measures the cumulative impact of a
scientist, which comes mainly from the quality of the research and
also from the size of the community in which the scientist
publishes.

In this paper I give an interesting twist on the \textit{h} index
and apply it further to the case of interesting topics and
compounds. It is common practice today that many so called
buzzwords are in use, in which a new person to the field hears the
word often but is left wondering, even what is meant by it,
without much explanation about the meaning being given. It is
getting increasingly difficult for a newcomer, i.e. PhD student or
undergraduate student to unravel all this information in order to
choose a masters or PhD topic that could be interesting and
relevant for their interest and future development. Especially in
solid state physics, there is much use of ``compound culture"
where a given student works on a single compound during a masters
or PhD course. A student or for that matter also a post-doc could
have the following questions: How much has already been done in
this topic or compound? Is it a so called ``hot topic"? Will many
people be interested in the results that I obtain from my thesis
or research? In today's wealth of information this could be a
difficult question to answer without doing a huge amount of
initial searching in the literature. A simple and quick, but by no
means full answer could be easily obtained by comparing the
\textit{h} index for topics or compounds, which will from now on
be called the ``\textit{h-b} index".

No rigorous mathematical treatment will be given here, but it will
follow from the procedures given from ref (1), in which the
\textit{h} index is defined, which I now assign as the
\textit{h-b} index. The following linear relation of the
\textit{h} index with the number of years from the first paper
published (eq. 6 from ref (1)) is given as

\begin{equation}
h \sim mn \label{eq:hbyears}
\end{equation}

With \textit{h}, substituted for \textit{h-b}. In this case, it is
assumed that \textit{h-b} increases linearly with the number of
years, \textit{n}. The gradient is thus given by \textit{m}, which
will vary from topic to topic or compound to compound. This linear
relation may or may not be the case, but it is an approximation
for which in a large number of cases will hold. In the case of a
scientist's name, this can be shown that from the first paper
published until a given maximum, in which afterwards the
\textit{h} index is constant. A plot of an authors \textit{h}
index against number of years (\textit{n}) will give an
approximate straight line\footnote{W. Marx, \textit{to be
published}.}. This also holds for a compound or topic search where
it is generally found that the \textit{hb} index increases
linearly with the number of years since the initial discovery,
until a given maximum when it is approximately constant. Even for
new expanding topics this initial linear behavior is found.

The \textit{h-b} index is found by entering a topic or compound in
the Thomson ISI Web of Science database and then ordering the
results in terms of citations, by largest first. The \textit{h-b}
index is defined as above in the same manner as the \textit{h}
index.

The number of years, \textit{n}, is found by the year of the first
paper published on this topic or compound until 2005. In this
manner, \textit{m} is easily calculated as the ratio of \textit{h}
to \textit{n}. Before any results or observations are given this
procedure will be discussed in terms of its legitimacy. The
database used, will in the case of topics or compounds, search the
title, keywords (incl. keywords plus) and the
abstract\footnote{Using Thomson ISI WoS, the abstracts are only
given for papers published after 1990.}. For example, for a given
compound, the paper could be about a high \textit{T}$_{c}$
superconductor, but include MgB$_{2}$ in its abstract, in which
case this paper would also be included for an MgB$_{2}$ search.
Although this hinders the results for an MgB$_{2}$ search if the
paper is highly referenced, this difficulty would occur in a
limited number of cases and would not affect the result in a large
way. As it is assumed that the title and keywords would contain
what the paper is written for. The abstract is the only way a
result could be ambiguous, but for the large majority of cases,
the results will hold. In the case of compounds, searching a
compound by its chemical name or formula should in a wide range
produce sound results, as the compound name is largely unique, as
in the same way the author's name. However, in some limited cases
another compound formula which contains the searched formula
string could be present. This is analogous to the case of an
author name that may not be unique, but for the most cases is (for
example, about 76\% of the Nobel prize winners in the last 20
years had unique names, from ref (1)). In the case of topics, the
search is more difficult and care must be taken. In the case,
where I specifically mention a topic (or compound), the exact
search string will be presented e.g. ``Kondo AND lattice". The
topic itself is not as concise as searching for compounds, due to
the more chance of the occurrence of the word rather than a
compound formula. As stated above there are some pitfalls, however
the results are nevertheless a good ballpark estimate, of the
current interest and the amount of people working in the area, now
and over the years. Next, I will give some results of searches I
have undertaken, and then discuss possible implications in respect
of the above questions and searches.

First is the discussion of compounds. The results I will present
are to compounds that were mostly unambiguously identified, with
no or very little ``washing" out of the \textit{h-b} index. In
some cases, e.g., H$_{2}$O, this string exists in many other
chemical formulas which contains ``H$_{2}$O", and thus appears
many times, so it is impossible to know what the number is for
only ``H$_{2}$O". Most compounds do not exhibit this feature, and
some of them are shown in table~\ref{tb:compounds}. It should be
noted that in calculating \textit{m}, the year of the first paper
about the compound was used, which in most cases is the paper of
synthesis and/or describing the crystal structure. This I believe
is nevertheless the best way of representing \textit{m}, of course
the \textit{h-b} index is regardless of this point.

It could be argued that using \textit{m} is not a good estimate of
how much interest has gone into a given compound, due to the fact
that the starting year I chose is probably the year of synthesis
or structure determination. There could be an initial time lag
where no further experiments were carried out on the compound
until a much later date. An interesting case could be MgB$_{2}$,
after the discovery of superconductivity, the work into this
compound exploded, but the time between the first paper in 1954,
gives a 40 year difference, this naturally affects its \textit{m}
number a great deal. In such cases a comparison of the
\textit{h-b} index is necessary. However, it is not the scope of
the paper to argue what is a good starting point for the
determination of \textit{m}. Some simple conclusions can be
pointed out by looking at table~\ref{tb:compounds}. The
superconductors (e.g., V$_{3}$Si, Nb$_{3}$Sn, MgB$_{2}$) all have
large values of \textit{h-b} and \textit{m}, a clear indication of
the enthusiasm and large number of scientists working in the field
of superconductivity. A comparison with research in rare earth
systems (e.g., PrPb$_{3}$, TmGa$_{3}$) shows that the number of
people working in these compounds is much less than
superconductivity (exceptions do exist, e.g., CeCu$_{2}$Si$_{2}$).
Work in C-60 bucky balls, is the largest that I have found to
date, with an \textit{m} of 5.2 which represents this as a unique
compound, in which there has been a large amount of work done.

\begin{table}
\caption{\textit{h-b} index for specific compounds given by a name
or chemical formula, sorted by ascending \textit{m}.
\label{tb:compounds}}
\begin{ruledtabular}
\begin{tabular}{|c|c|c|}

Compound & \textit{h-b} index  &  m   \\
\hline PrPb$_{3}$ & 6 & 0.26 \\
\hline TmGa$_{3}$ & 6 & 0.30 \\
\hline Si28 & 17 & 0.31 \\
\hline CeB$_{6}$ & 32 & 0.76 \\
\hline V$_{3}$Si & 39 & 0.77 \\
\hline Ni$_{2}$MnGa & 37 & 0.82 \\
\hline Nb$_{3}$Sn & 48 & 0.94 \\
\hline MgB$_{2}$ & 67 & 1.31 \\
\hline CeCu$_{2}$Si$_{2}$ & 39 & 1.44 \\
\hline SrTiO$_{3}$ & 94 & 1.96 \\
\hline GaN & 144 & 2.12 \\
\hline C-60 & 182 & 5.20 \\

\end{tabular}
\end{ruledtabular}
\end{table}

Therefore I find from looking at all my results from searches
carried out on compounds the following conclusions.

i) 0 $<$ \textit{m} $\leq$ 0.5, i.e. a maximum \textit{h-b} of 20
after 40 years. This represents a compound which is likely to be
of interest to the researchers in that particular field of
research, where the field is a smaller community.

ii) 0.5 $\leq$ \textit{m} $\leq$ 2, i.e a maximum \textit{h-b} of
40 after 20 years. This represents a compound that is likely to be
a hot topic area of research, where the community is very large.
Or a compound with very interesting features.

iii) \textit{m} $\geq$ 2, i.e a minimum of \textit{h-b} of 40
after 20 years. This reflects a unique compound, which has far
reaching consequences rather than just in its own area of
research. It is likely to be a compound with application purposes,
or unique characteristic features.

Next, I will discuss the larger area of topics, which of course
extends much further than just the scope of solid state physics.
The searching of topics was performed in the same manner as with
the compounds above. There are a few points of interest which
should be addressed first. Searching a topic would give a hit,
when the string is represented in the title, keywords or the
abstract. This is expected in the majority of cases to be
legitimate. If a false hit is present, this would occur in most
cases from the abstract, but its effect is neglected here. Some
searches of certain topics is meaningless, e.g. ``magnetism" or
``specific heat", and would give answers in which not much
information could be extracted. However for today's so called
``hot topics" such searches could be used. Table II shows some of
the search results.

In the same way as with specific compounds, the use of certain
keywords describing the topic, may take some time after the
initial discovery to surface in common use, so there may also be
some initial time lag, between the true discovery, and the naming
of it. However in most cases this would only be a year at the
most. Table~\ref{tb:topics} shows the results of some searches for
specific topics in many areas of physics. Some of the older topics
are most likely to be found at the top of the table (as the table
is \textit{m} ascending), which is likely also caused by the
increasing number of researchers in all fields of science over the
years. Therefore, most of the new hot topics are represented at
the bottom of the table, e.g., nanowires. In some cases it may be
more useful in order to compare by \textit{h-b} number, which of
course is still modified by the total number of people or groups
in that area, but not dependant on \textit{n}, the number of
years. Some more general conclusions are now given from the
searches of topics:

\begin{table}
\caption{\textit{h-b} index for specific topics, search strings
are given, sorted by ascending \textit{m}. \label{tb:topics}}
\begin{ruledtabular}
\begin{tabular}{|c|c|c|}

Topic & \textit{h-b} index  &  m   \\
\hline Borides & 46 & 0.44  \\
\hline pyrochlore & 61 & 0.62 \\
\hline Spin flop & 34 & 0.83 \\
\hline Optical lattice & 43 & 0.90 \\
\hline Antiferroquadrupolar & 18 & 1.00 \\
\hline amorphous silicon & 116 & 1.10 \\
\hline Spin frustration & 30 & 1.36 \\
\hline ferroelectricity & 78 & 1.39 \\
\hline Spin liquid & 45 & 1.55 \\
\hline kondo AND lattice & 63 & 1.97 \\
\hline perovskites & 103 & 2.10 \\
\hline spin ice & 17 &  2.13 \\
\hline magnetoresistance & 172 & 2.39 \\
\hline quantum information & 65 & 2.41\\
\hline geometrical frustration & 21 &  2.63 \\
\hline quantum critical point & 42 & 2.63 \\
\hline porous silicon & 104 & 3.25 \\
\hline spin glass & 108 &  3.38 \\
\hline Spin valve & 48 & 3.43 \\
\hline heavy fermion & 97 & 3.73 \\
\hline superstrings & 99 &  3.96 \\
\hline Teleportation & 61 &  5.08 \\
\hline quantum computation & 73 & 5.21 \\
\hline M-theory & 79 & 6.58 \\
\hline giant magnetoresistance & 116 &  6.82 \\
\hline fullerenes & 140 & 7.78 \\
\hline quantum dots & 149 & 7.84 \\
\hline Nanowires & 105 &  8.75 \\
\hline carbon nanotubes & 167 & 12.85 \\
\end{tabular}
\end{ruledtabular}
\end{table}

i) A large \textit{m} (\textit{m} $\geq$ 3) number is likely to
come from a topic which is a so-called ``hot topic", as a large
\textit{m} number comes from the fact that the number of years is
small, while maintaining a large enough \textit{h-b} number.

ii) A large \textit{h-b} number (\textit{h-b} $\geq$ 100) and a
large \textit{m} number (\textit{m} $\geq$ 3) represents a topic
in which there has been a lot of research already, which was and
still is a hot topic.

iii) A small \textit{m} (\textit{m} $\leq$ 2) but a large
\textit{h-b} number (\textit{h-b} $\geq$ 100) probably reflects
this as a older topic, which has a good contribution throughout
the years, which is why the \textit{m} number is low.

Of course these rules are by no means complete, but kind of a
first approach, that one could follow, but not adhere too. There
are some topics which are revolutionary in some way, like carbon
nanotubes or nanowires, which have had a huge amount of work done,
in a very short amount of time.

I have presented here mainly the results for the case of physics,
and with a little more emphasis on solid state physics. I am sure
that this would also work for any area of science, where topics
can be clearly separated from each other. I have discussed a lot
of topics and compounds, without talking about the physics itself.
My results are only based on a few numbers. As for the case of a
scientist, a number does not reflect the potential of a person. In
this case, also the number does not reflect the science or the
interesting physics. Rather it does reflect the interest in the
immediate community and further, and the amount of work which has
been done. It should be noted that using this method in order to
join a particular area of research is another question and based
on many other variables. However, this method can be used to get a
feeling of what topic or compound is mainstream research at this
present time. Therefore, there can be some conclusions drawn
without going into the details of the specific research area, this
I find itself is a success of this technique. With regard to the
questions that I posed at the start of the article. I would now
like to answer or attempt to reason why this technique could help
a student or researcher find his way through topics. Regarding the
question of finding out how much research has gone into the topic
already, I think this is a useful way of finding it out, and
probably represents the best way in order to get a feeling of how
much research has already been done. The question of whether this
is a ``hot topic" or not is difficult to answer in a number, as
``hot topics" come from interesting physics, or interesting
features. However, what can be immediately concluded is that a
large number of people work in the area, which in itself is a way
to identify it as a ``hot topic". In a similar manner working in a
``hot topic" area will make your work more visible by a larger
community.

Concerning Hirsch's comments regarding membership of a national
academy or institution by comparing a given researchers \textit{h}
index. An interesting point could be made by looking at the
\textit{h-b} index when allocating funding money from grants. This
method could be used to compare different researchers applying for
a given research grant in the same field. By searching for a
researchers name and topic of proposed research, then comparing
the \textit{h-b} index, could give interesting results and
comparative features of already published work in the area.

In summary, I have proposed a tool to compare different topics and
compounds based on the result obtained by J. E.
Hirsch\cite{jeh:hirsch,jeh:preprint}. I have found interesting
correlations, which can be used to give some conclusions which
point to whether this is an ``older topic" or a new ``hot topic"
by comparing the \textit{h-b} index and \textit{m}. It has been
shown that this method could help new comers to the field, to
identify how much interest and work has already been achieved in
their chosen area of research.


\begin{acknowledgments}
I thank all the colleagues at the Max-Planck-Institute for a
stimulating atmosphere and discussing and criticizing the idea in
a concise way and giving me enough enthusiasm in order to research
the idea and publish it.
\end{acknowledgments}

Note added in proof: Due to the recent interest and published
articles on this idea,\cite{nature,physics} it should be noted
that I offer a technique to compare selected topics, but do not
offer an overall ranking of topics for an entire discipline in
this manuscript.

\bibliography{hb}

\end{document}